\documentclass[modern]{aastex63}

\usepackage{graphicx}
\usepackage[suffix=]{epstopdf}
\usepackage{natbib}
\usepackage{amsmath}
\usepackage{url}
\usepackage{xspace}

\providecommand{\e}[1]{\ensuremath{\times 10^{#1}}}

\newcommand{\Kepler}{\textsl{Kepler}\xspace}

\begin{document}

\title{10 Years of Stellar Activity for GJ 1243}

\correspondingauthor{James. R. A. Davenport}
\email{jrad@uw.edu}

\author{James. R. A. Davenport}
\affiliation{Astronomy Department, University of Washington, Box 951580, Seattle, WA 98195, USA}

\author{Guadalupe Tovar Mendoza}
\affiliation{Astronomy Department, University of Washington, Box 951580, Seattle, WA 98195, USA}
\affiliation{Astrobiology Program, University of Washington, 3910 15th Ave. NE, Box 351580, Seattle, WA 98195, USA}

\author{Suzanne L. Hawley}
\affiliation{Astronomy Department, University of Washington, Box 951580, Seattle, WA 98195, USA}

\begin{abstract}
The flaring M4 dwarf GJ 1243 has become a benchmark for studying stellar flare and starspot activity thanks to the exceptional photometric monitoring archive from the \Kepler mission. 
New light curves from the TESS mission for this star allow precise stellar activity characterization over more than a decade timescale.
We have carried out the first flare and starspot analysis of GJ 1243 from over 50 days of data from TESS Sectors 14 and 15. 
Using 133 flare events detected in the 2-minute cadence TESS data, we compare the cumulative flare frequency distributions, and find the flare activity for GJ 1243 is unchanged between the \Kepler and TESS epochs.
Two distinct starspot groups are found in the TESS data, with the primary spot having the same rotational period and phase as seen in \Kepler. The phase of the secondary spot feature is consistent with the predicted location of the secondary starspot and measurement of weak differential rotation, suggesting this secondary spot may be long-lived and stable in both latitude and longitude.
As expected for this highly active star, the constant spot and flare activity reveal no sign of solar-like activity cycles over 10 years. However, we highlight the unique ability for \Kepler and TESS to use flare rates to detect activity cycles.
\end{abstract}

\section{Introduction}

The rapidly rotating, magnetically active, M4 star GJ 1243 is perhaps one of the best characterized flare stars over the past decade. Four years of nearly continuous 30-minute cadence observations by the \Kepler mission \citep{borucki2010} found prominent starspots for GJ 1243 (\Kepler ID 9726699) rotating with a period of 0.59 days \citep{savanov2011}, and with indications of very slow starspot evolution due to weak differential rotation \citep{davenport2015b,giles2017}. \Kepler also obtained 11 months of ``short cadence'' (1-minute) monitoring, which were ideal for characterizing the flare activity for this star \citep{hawley2014, silverberg2016}, amassing the largest catalog of homogeneously observed flares for any star besides the Sun \citep{davenport2014b}. During the \Kepler original mission,  GJ 1243 was observed for four years, from May 2009 until May 2013. As the brightest single flaring M dwarf in the \Kepler field, GJ 1243 has become the benchmark flare star studied by exoplanet-searching telescopes. Continuing to measure the flares and starspots from this important \Kepler star will allow us to characterize the long-term behavior of surface magnetic activity for fully-convective stars, as has been done for very few stars in the past \citep[e.g. V374 Peg][]{vida2016}.

The Transiting Exoplanet Survey Satellite \citep[hereafter TESS][]{Ricker2014}, in many ways the successor to \Kepler, has recently revisited GJ 1243 (TESS ID 273589987). Short-cadence (2-minute) light curves for GJ 1243 are available from Sectors 14 and 15 (July 18 2019 through September 11 2019), providing the most detailed light curve for this star since the end of the \Kepler mission, and giving a 10-year total observing baseline since the first \Kepler data for this star became available (30-minute cadence data from \Kepler Quarter 0 in 2009). 

The decade timescale of precision photometric monitoring using the combined \Kepler--TESS data enables a host of new stellar activity studies. Most notably, this provides stellar activity estimations (e.g. starspot amplitudes and rotation rates, and flare activity levels) on a timescale that is comparable to the Sun's $\sim$11 yr activity cycle. As \citet{scoggins2019} has recently highlighted, variations in flare rates determined with missions like \Kepler and TESS are a promising method for detecting stellar activity cycles, since the flare rate for the Sun is observed to vary by roughly an order of magnitude between solar maximum and minimum \citep[e.g.][]{veronig2002,aschwanden2012}. 
Long term monitoring of starspot modulations and rotation periods may also provide indications of stellar magnetic activity cycles \citep[e.g.][]{messina2002,nielsen2019,morris2019}, or differential rotation \citep{reinhold2013}. 
While rapidly rotating M dwarfs like GJ 1243 and V374 Peg are not expected to show solar-like activity cycles, the variance of their surface magnetic activity over longer timescales -- especially in precision estimates such as flare rates using space-based photometry -- has been relatively unexplored.

Here we explore the starspot and flare activity for GJ 1243 using a combination of light curves from \Kepler and TESS that span a 10-year baseline. As we do not have continuous monitoring over this 10-year baseline, the \Kepler and TESS data instead provide point-in-time measurements of the flare and starspot activity for GJ 1243. We introduce the TESS short-cadence data in \S\ref{sec:data}. In \S\ref{sec:flare} we define our sample of flares from TESS, and compare the flare rate with the benchmark from \Kepler. As in \Kepler, two distinct starspots regions are found in the TESS light curves, and we explore their evolution over the past decade in \S\ref{sec:spots}. Finally we conclude with a look ahead at the promising future for stellar activity studies combining \Kepler and TESS data in \S\ref{sec:discuss}, and provide a brief summary of our results in \S\ref{sec:sum}.

\begin{figure*}[!t]
\centering
\includegraphics[width=6in]{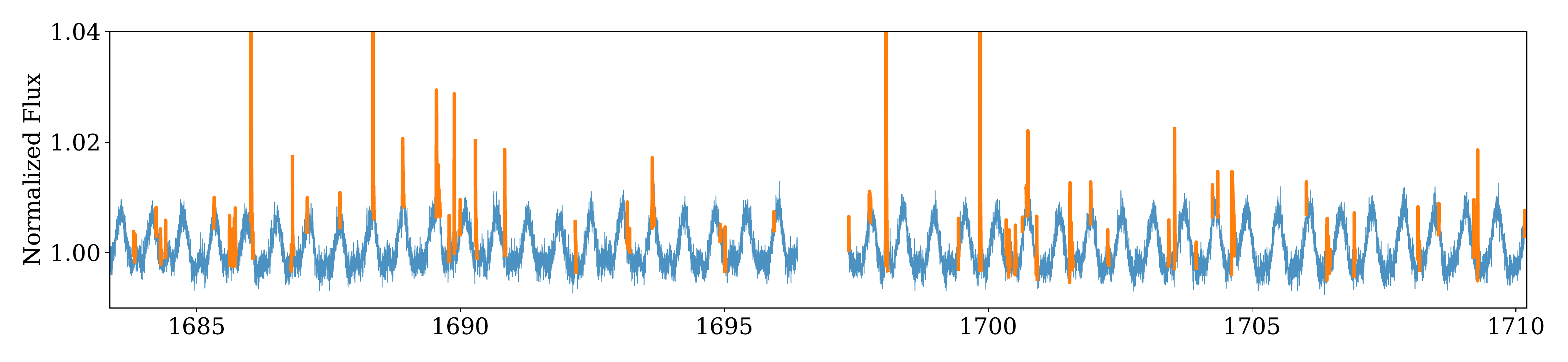}\\
\includegraphics[width=6in]{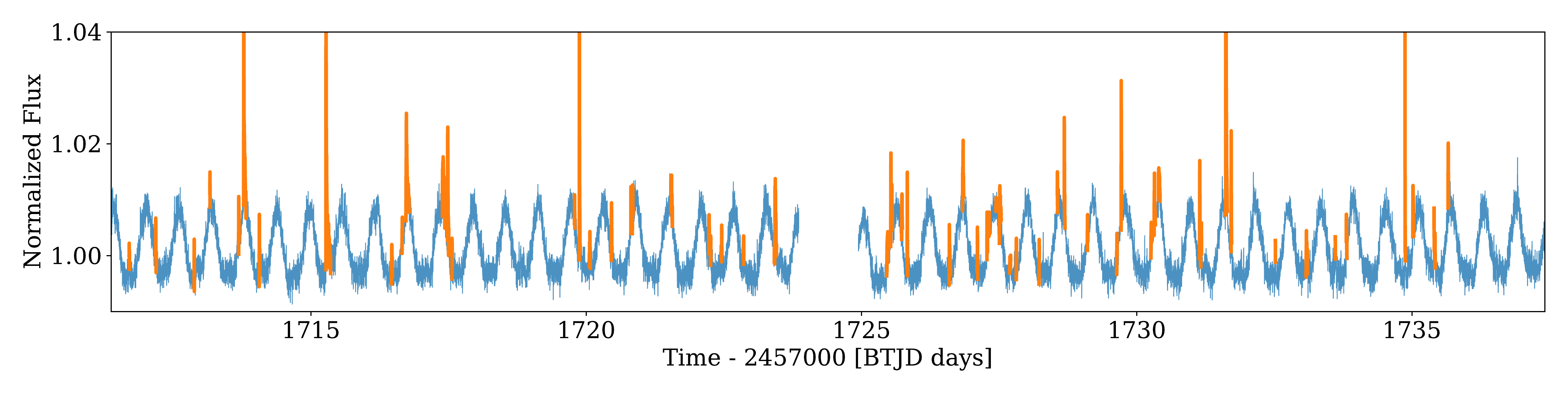}
\caption{
TESS Sector 14 and 15 light curves for the rapidly rotating, flaring M dwarf, GJ 1243. Using {\tt FBEYE} we identified 133 flare events (orange) over the 54 days of 2-minute cadence observations. For visual clarity we have not shown the full range of the largest detected flares, which reach amplitudes of 23\%. 
}
\label{fig:lc}
\end{figure*}

\section{TESS Short Cadence Data}
\label{sec:data}

TESS is a space-based, exoplanet-searching telescope that was launched in 2018. Over the course of the primary two-year mission, TESS will survey $\sim$85\% of the entire night sky with at least 27 days of continuous observation in search of transiting exoplanets around the brightest nearby stars \citet{Ricker2014}. 
To date, TESS has confirmed 45 new exoplanets and identified more than a thousand objects of interest\footnote{As of 2020 April,  https://tess.mit.edu/publications/}, while simultaneously studying the astrophysics of more than 200,000 stars with short-cadence (2-minute) data. 
The data from TESS Sectors 14 and 15 from mid-2019 (and again partially in mid-2020 with Sector 26) provide a unique opportunity to revisit the original \Kepler field. By combining the data from both missions we can characterize stellar activity for GJ 1243 on a decade timescale\footnote{\Kepler Quarter 0 was observed in May 2009}. 

We use the short cadence data from sectors 14 and 15, totaling 50.47 days of 2-minute observations. We are using the {\tt PDCSAP\_FLUX} light curves provided by the TESS mission directly, and require the {\tt Quality} flag be set to 0 to ensure the highest fidelity data possible. Light curves for both Sector 14 and 15 are shown in Figure \ref{fig:lc}, with our detected flares (described below) highlighted.

\section{Comparing Flare Activity}
\label{sec:flare}

To gather a sample of flares from the TESS data, we examined the short cadence (2-minute) light curve for Sectors 14 and 15 using {\tt FBEYE} \citep{davenport2014b}. This flare finding suite was originally designed to manually build a pristine flare sample for GJ 1243 with \Kepler. 
In {\tt FBEYE} the light curves are first smoothed with an iterative three-pass approach using a variable sized smoothing kernel based on \citet{supersmoother}. This is designed to remove slow variability primarily from the starspots, and preserve the flares. Candidate flare epochs with fluxes greater than 2.5$\sigma$ above the smoothed light curve are then highlighted for the user. As \citet{davenport2014b} illustrate, while this automated step reliably detects the peaks of large flares, it is not optimized to find all small events, nor accurately separate the graduate decay phase of the flare from the underlying starspot variability. A by-eye validation of the entire 50.47 day light curve is then performed in {\tt FBEYE}, which recovered a total of 133 flare events.

\begin{figure}[!t]
\centering
\includegraphics[width=4in]{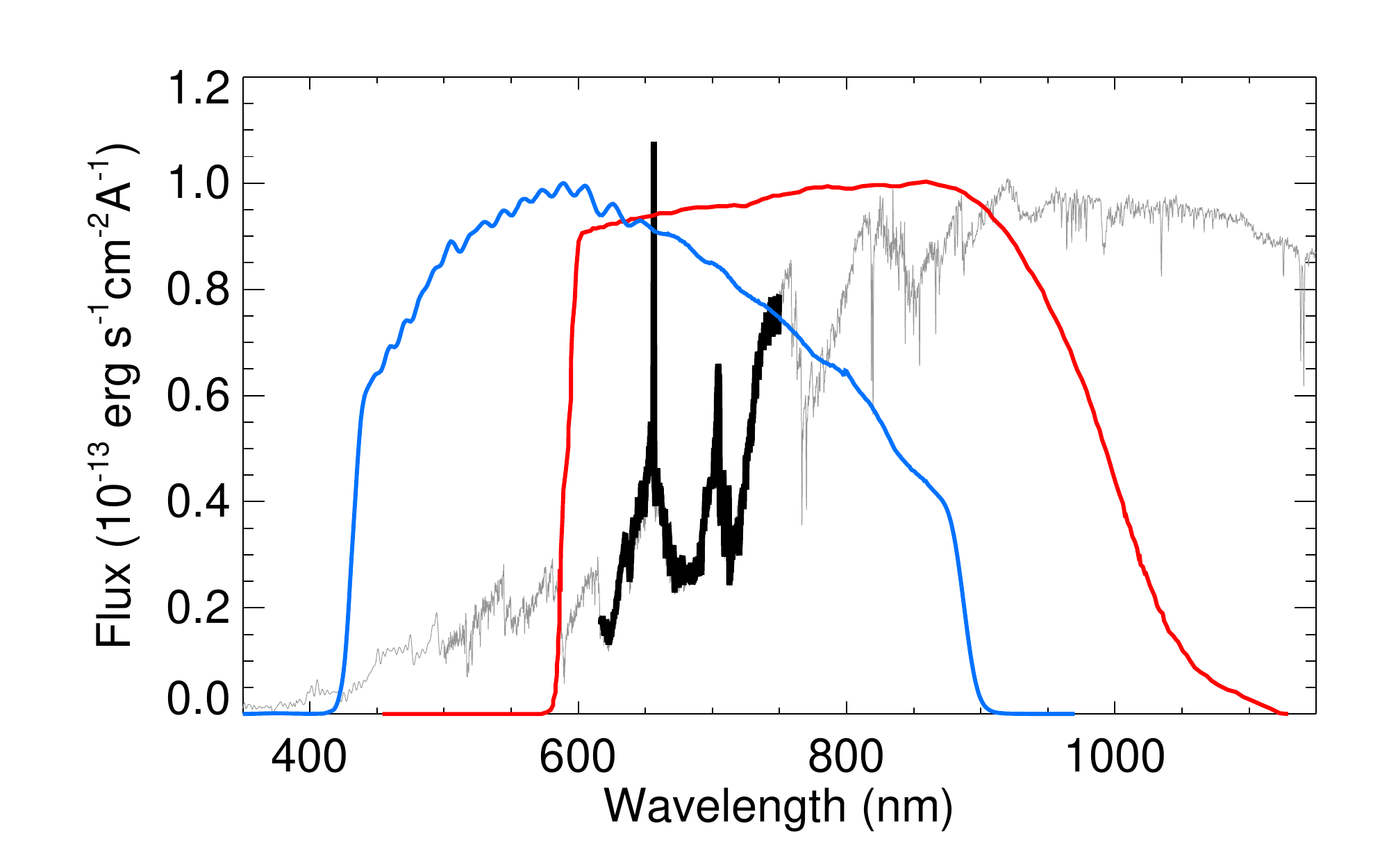}
\caption{
Flux calibrated spectrum for GJ 1243 (thick black line) from \citet{pmsu1}, with scaled NUV--NIR M4 template (thin grey line) from \citet{davenport2012}, which is used to calculate the quiescent luminosity for GJ 1243 in the \Kepler (blue line) and TESS (red line) bandpasses.
}
\label{fig:spec}
\end{figure}

Rather than flare energies, {\tt FBEYE} provides the {\it equivalent duration} for each event, defined as the integral of the flare in zero-registered normalized flux, which has units of seconds \citep[see][]{huntwalker2012}. The energy for each flare is then determined by multiplying the equivalent duration by the star's luminosity. 
Following \citet{davenport2016}, we determined the luminosity for GJ 1243 in both the \Kepler and TESS bandpasses. The \Kepler and TESS absolute magnitudes were estimated by matching the $g-J$ color of GJ 1243 to a 1 Gyr, solar metallicity PARSEC isochrone \citep{bressan2012}. We explored isochrones with ages up to 5 Gyr, and they did not significantly change change the resulting luminosity estimates for GJ 1243.
These AB magnitudes were converted to fluxes, and set to a distance of 10 pc to determine the luminosity of the star in each filter.  
We find $\log L_{Kep}=30.68\pm0.04$ erg s$^{-1}$, and $\log L_{TESS}=31.06\pm0.04$ erg s$^{-1}$. 
Uncertainties in the luminosity were estimated numerically, by re-sampling the observed $g-J$ color 1000 times using a Gaussian kernel with a standard deviation equal to the observed $g-J$ errors, and finding the standard deviation of the resulting luminosity estimates from the isochrone. 
The \Kepler luminosity used here is very close to the $\log L_{Kep}=30.67$ erg s$^{-1}$ that \citet{hawley2014} found by convolving spectrophotometric data from \citet{kowalski2013} with the \Kepler bandpass. 
\citet{hawley2014} claim a conservative uncertainty on $\log L_{Kep}$ of $\pm 0.2$, but noted that when using flux calibrated spectral observations from possibly non-photometric conditions they get estimates on the luminosity that were 0.15--0.7 dex lower.
We also attempted to use flux calibrated refernece spectra to estimate the luminosity for GJ 1243 in both the \Kepler and TESS bandpasses. 
As shown in Figure \ref{fig:spec}, each filter curve was convolved with a M4 NUV--NIR spectral template from \citet{davenport2012}, which was normalized to the flux-calibrated optical spectrum for GJ 1243 from \citet{pmsu1}. We find $\log L_{Kep}=30.04$ erg s$^{-1}$, and $\log L_{TESS}=30.35$ erg s$^{-1}$ -- about 0.6 dex lower than our isochrone estimates above, but with a comparable difference in luminosity between the filters. 
\citet{pmsu1} also note their spectral library is observed under potentially non-photometric conditions, and so our spectro-photometric luminosity estimate here is also consistent with the results from of \citet{hawley2014}. We emphasize the specific luminosity value used for the \Kepler bandpass does not impact our results in comparing the flare activity {\it between} \Kepler and TESS for GJ 1243, since we have estimated the luminosity in both bands consistently. Rather, we urge caution when comparing specific flare rates between studies that use differing methods for determining stellar luminosity (e.g. spectrophotometry versus isochrone or model fits), since this can result in large ($>$0.5 dex) variations in the implied specific flare rate.

Since the \Kepler and TESS filter curves shown in Figure \ref{fig:spec} cover different wavelength regions, we also briefly explored how comparable the flare energies determined using these two filters would be. We convolved a 10,000 K blackbody curve with each filter, since this is a typical flare temperature assumed when estimating flare energies. Though the \Kepler filter is centered closer to the peak wavelength of this blackbody curve, the TESS filter's substantially wider wavelength coverage results in a 2.4\% {\it larger} flux response. Given typical uncertainties in the distance estimation and flux calibration for nearby stars, this indicates TESS and \Kepler have remarkably similar flare energy yields, and are well suited for comparison.

Flare frequency distribution (FFD) diagrams have been used to compare the occurrence rate of flares on stars as a function of flare energy \citep[e.g.][]{lme1976,shibayama2013}. This reverse sorted, cumulative flare distribution, characteristically shows many small energy events and very few high energy flares. Given their stochastic occurrence, FFDs help us understand the rate of flares over many  orders of magnitude in event energy. \citet{scoggins2019} used both FFDs and the integrated flare energy metric, $L_{fl}/L_{Kep}$ introduced by \citet{lurie2015}, to find flare activity variations within \Kepler data. $L_{fl}/L_{Kep}$ is a more robust measurement, as it averages over uncertainties in individual event energies, but assumes fixed observing conditions throughout the sample.
As our flare data for GJ 1234 comes from two telescope surveys with substantially different apertures and observing baselines, FFDs are the best parameter space for comparing our flare activity levels \citep[see \S3 of ][]{davenport2019}.

\begin{figure}[t]
\centering
\includegraphics[width=4in]{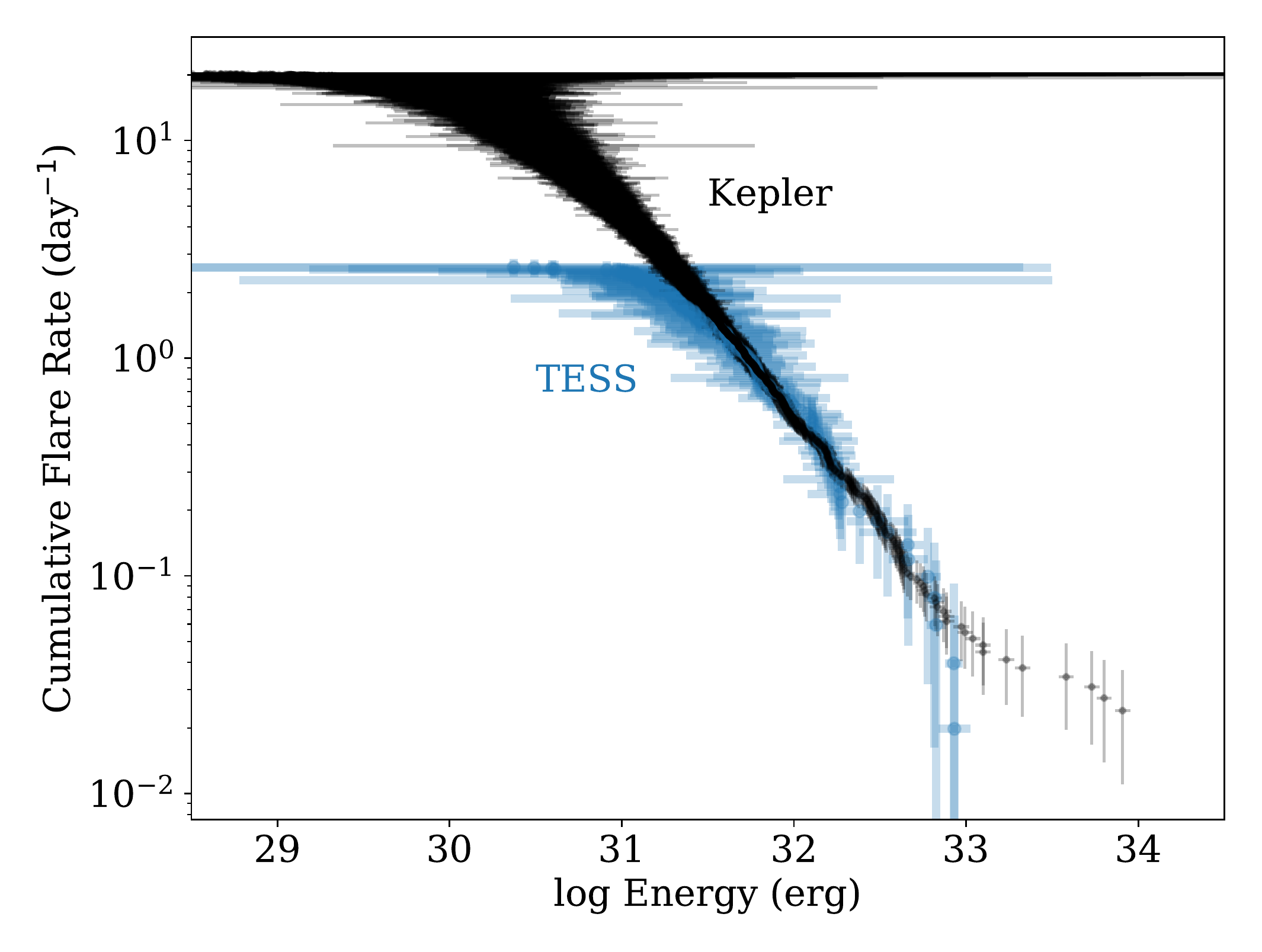}
\caption{
The cumulative flare frequency distribution (FFD) for the 6107 flares observed on GJ 1243 with \Kepler (black line) from \citet{davenport2014b}, and the 133 flares from TESS from this work (blue line). The flares are sorted from largest to smallest energy events per day. We can fit a power law to each data set (see section 3 for details). The turnover at low energies is due to lower detection efficiency of small flares \citep[e.g. see][]{paudel2018,ilin2019}.
}
\label{fig:ffd}
\end{figure}

Figure \ref{fig:ffd} shows the FFDs from the TESS and \Kepler observations. The differences in turn over rates between the two data sets is correlated with the amount of data collected. The FFD for the TESS data includes 50.47 days worth of 2-minute observations, while the \Kepler data spans 11 months short cadence, 1-minute monitoring. The apparently change in slope and rapidly increasing uncertainties at lower event energies are due to low signal-to-noise flares that have growing incompleteness in their recovery \citep{davenport2016}. For large energy flares we become limited by the observing duration, and are dominated by Poisson counting errors.

Error bars shown for the FFDs in Figure \ref{fig:ffd} were produced following the approach of \citet{davenport2016a} and \citet{howard2018}. The uncertainty in the cumulative flare rates (y-axis) were computed using the 1$\sigma$ confidence intervals for event counting statistics determined from the Poisson distribution by \citet{gehrels1986}. Their Eqn. 7 provides a simple analytic approximation to determine the uncertainty for counting a number of events (in this case flares) even for small numbers of events (e.g. for the rare large energy flares).
Uncertainty in the flare energies were computed using the same approach used in computing errors for spectral line equivalent widths, following Eqn. 6 from \citet{vollmann2006}. This uses the flare's total duration, the equivalent duration (analogous to the equivalent width), and the signal to noise ratio of the light curve. 
These energy errors have the expected behavior (i.e. errors decrease for higher energy events). Errors for the smallest flares cataloged increase dramatically as shown in Figure \ref{fig:ffd}, indicating these smallest events are statistically indistinguishable from noise. The combination of these occurrence rate and event energy uncertainties gives a characteristic ``bow-tie'' appearance to the FFD.

The FFD is typically modeled using a power law distribution, which has been shown to extend with a fixed slope over 10 orders of magnitude in flare energy for  Solar-type stars for "nanoflares" to "superflares" \citep{shibayama2013,maehara2015}. As noted in \citet{scoggins2019}, stars with significant differences in the power law {\it offset} (i.e the intercept in log-log space) would indicate changes in flare activity levels. A star whose flare activity level changed over a few years timescale may be undergoing activity cycles. We fit both the \Kepler and TESS data with a linear model in log-log space. Within the resulting error uncertainties we find the FFDs for both the TESS and \Kepler data are the same, revealing no sign of a stellar activity cycle over this decade. Our resulting power law fit for the TESS data was 
$\log_{10}(\nu) = -0.95 \pm 0.02 \times \log_{10} E + 29.4 \pm 0.7$,
and for the \Kepler FFD data was
$\log_{10}(\nu) = -0.942 \pm 0.001 \times \log_{10} E + 29.27 \pm 0.04$.
The larger uncertainties of the TESS data are to be expected, given the lower signal-to-noise of the light curves and the shorter duration of observation. However, it is encouraging that TESS can effectively characterize the ongoing flare activity level of \Kepler stars.

\section{Comparing Starspots}
\label{sec:spots}

Four years of nearly continuous monitoring for GJ 1243 from \Kepler revealed the presence of two distinct starspot features, which appear as slowly evolving quasi-sinusoidal modulations in the light curve (e.g. Figure \ref{fig:lc}). 
\citet{davenport2015b} produced a detailed tracing of the phase (or equivalently, longitude) of both starspot features over the entire \Kepler dataset. They noted the primary spot was approximately stable in both phase and amplitude, while the secondary feature evolved on a timescale of hundreds of days. Forward modeling of the \Kepler light curve suggested the primary spot was located at a high latitude, while the secondary spot was at a lower latitude.
Interestingly, during the final two years of the \Kepler data, the secondary spot appeared to evolve almost monotonically in phase, which \citet{davenport2015b} interpreted as the signature of differential rotation of a long-lived spot feature.

The combined \Kepler--TESS 10-year light curve allows us to improve the rotation period estimation for GJ 1243. Using a Lomb-Scargle Periodogram, we measured the rotation period to be $P_{rot} = 0.5925974$ days, 1.2\e{-6} days longer than the estimate from \citet{davenport2015b}. 
To estimate an uncertainty on this period measurement, we use the analytic approximation developed by \citet{mighell2013} for \Kepler eclipsing binaries. Scaling their expression with the 10-year baseline of observations between \Kepler and TESS (e.g. following their Figure 5), we estimate an uncertainty of $\sigma_P$ = 1\e{-7} days. Similarly, the \citet{davenport2015b} period from 4 years of \Kepler data had an uncertainty of $\sigma_P$ = 4\e{-7} days. We note that the errors approximations from \citet{mighell2013} assume a truly constant periodic source and continuous data, and are therefore likely underestimated. We consider the difference between the rotation period for GJ 1243 from \citet{davenport2015b} and this current work to therefore be negligible ($\sim$2.8$\sigma$), but report the updated period estimate here for the benefit of future studies.

\begin{figure*}[!t]
\centering
\includegraphics[width=6in]{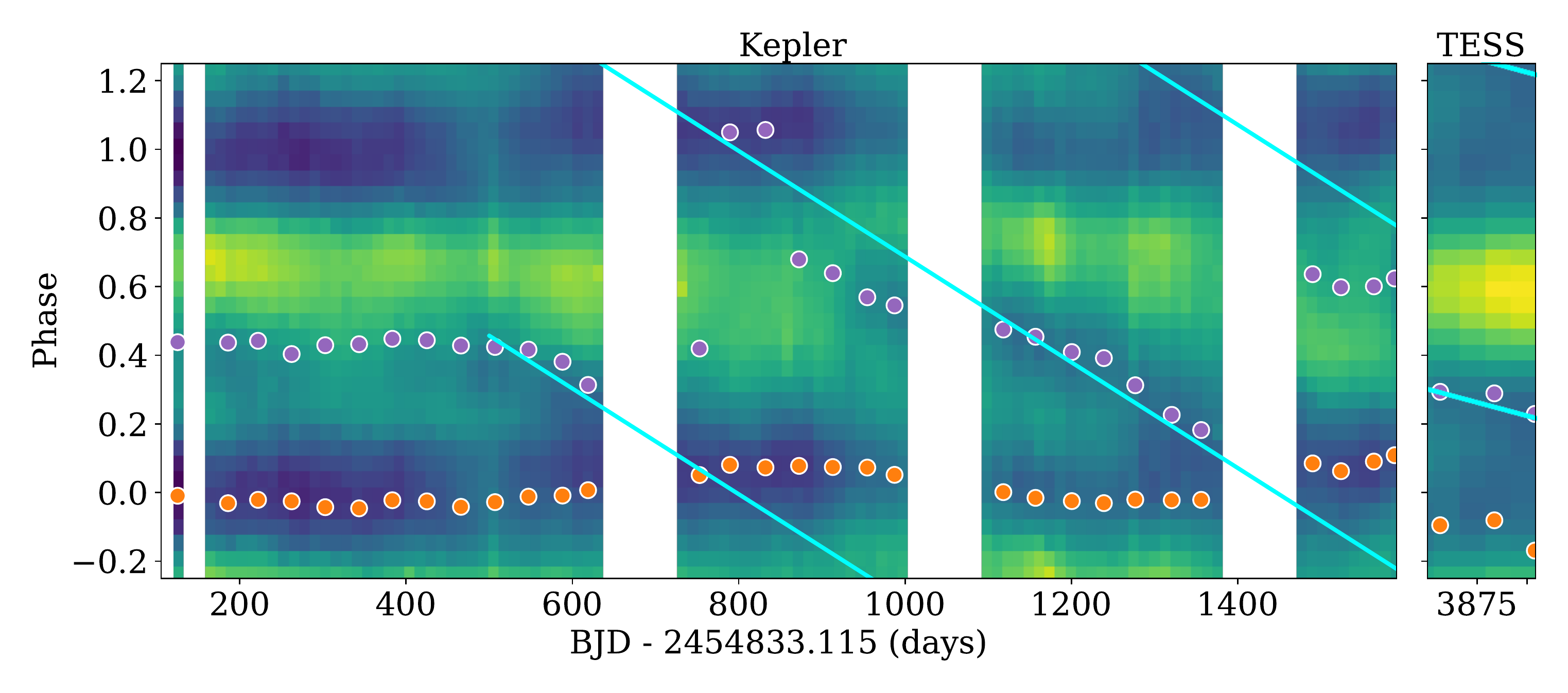}
\caption{
Mapping of relative flux (pixel shade, from dark to light) as a function of rotation phase and time for both the \Kepler and TESS data of flux vs time and phase. The dark band centered at Phase=1 is the primary starspot group, which remains essentially unchanged over the entire 10 year dataset. Gaussian fits to the phase curves are used to trace the locations of the primary and secondary starspot in 40-day bins (orange and purple points). The differential rotation of the secondary starspot from \citet{davenport2015b} is fit from Time = 400 to 1400 days (blue line). Projecting forward to the TESS observations, the secondary spot appears in phase with this simple prediction, which is likely a coincidence. 
}
\label{fig:map}
\end{figure*}

In Figure \ref{fig:map} we reproduce the phase versus time evolution map of flux from \citet{davenport2015b} for the 4-year \Kepler long cadence (30-minute) data\footnote{Here we used the final Data Release 25 version of the long cadence \Kepler data}, as well as the new TESS Sectors 14 and 15 short cadence data. All data have been put onto the \Kepler Barycentric Julian Date time. 
The \Kepler long-cadence data has been normalized by the per-Quarter median flux value using the {\tt Lightkurve} package. 
Pixel color and shade encode the flux of the star as a function of time and rotational phase. 
The primary starspot is seen as a dark band (i.e. lower normalized flux in Figure \ref{fig:lc}) centered at Phase=0. Note we have increased the relative flux response of the TESS data here by a factor of 2 to approximately match the starspot amplitude seen in \Kepler. This difference in spot amplitude is expected due to the redder wavelength center of the TESS filter (e.g. Figure \ref{fig:spec}), which leads to a lower spot-to-star contrast ratio (i.e. spots are higher amplitude at bluer wavelengths).

To analytically trace the positions of the starspots over the 10-year baseline, we have split the combined \Kepler--TESS light curve into 40 day bins of time, and fit the phase-folded data within each bin with two Gaussian curves. The center phase of the primary and secondary spots are shown as circles in Figure \ref{fig:map}. Modeling with Gaussian curves can cause blending and result in lower velocity amplitudes (this is often the case when decomposing multiple radial velocity peaks in spectroscopy). The same effect can be seen here, with the two peaks blending together, causing the primary starspot position to ``wiggle'' in phase due to blending with the secondary spot.

\citet{davenport2015b} estimated the surface shear due to differential rotation by measuring a monotonic change in phase of the secondary starspot relative to the primary. Specifically, they focused on the clear evolution in the secondary spot seen in Figure \ref{fig:map} from Time $\sim$950 to $\sim$1400 days. We note if this evolution is propagated {\it backwards} in time, the spot feature around Time$\sim$450 days appears to be in sync as well. However, as \citet{davenport2015b} found, the differential rotation evolution of this particular starspot does not correctly align with the secondary spot feature observed at Phase$\approx$0.5 in the first $\sim$300 days of the \Kepler data.

We updated the phase evolution estimate from \citet{davenport2015b} to include this earlier starspot, shown as a continuous blue line in Figure \ref{fig:map}. As \citet{davenport2015b} illustrate, the negative slope of this line indicates a spot moving {\it faster} than the rotation period used to phase-fold the data. The actual slope of the linear fit is the inverse of the rotation lap time: $1 / t_{lap} = $ -1.5393\e{-3}, which can be converted to a rotation shear of $\Delta \Omega  = 2\pi / t_{lap} = $ -0.00967 rad day$^{-1}$, somewhat stronger than measured previously by \citet{davenport2015b}.

\begin{figure}[!t]
\centering
\includegraphics[width=3.5in]{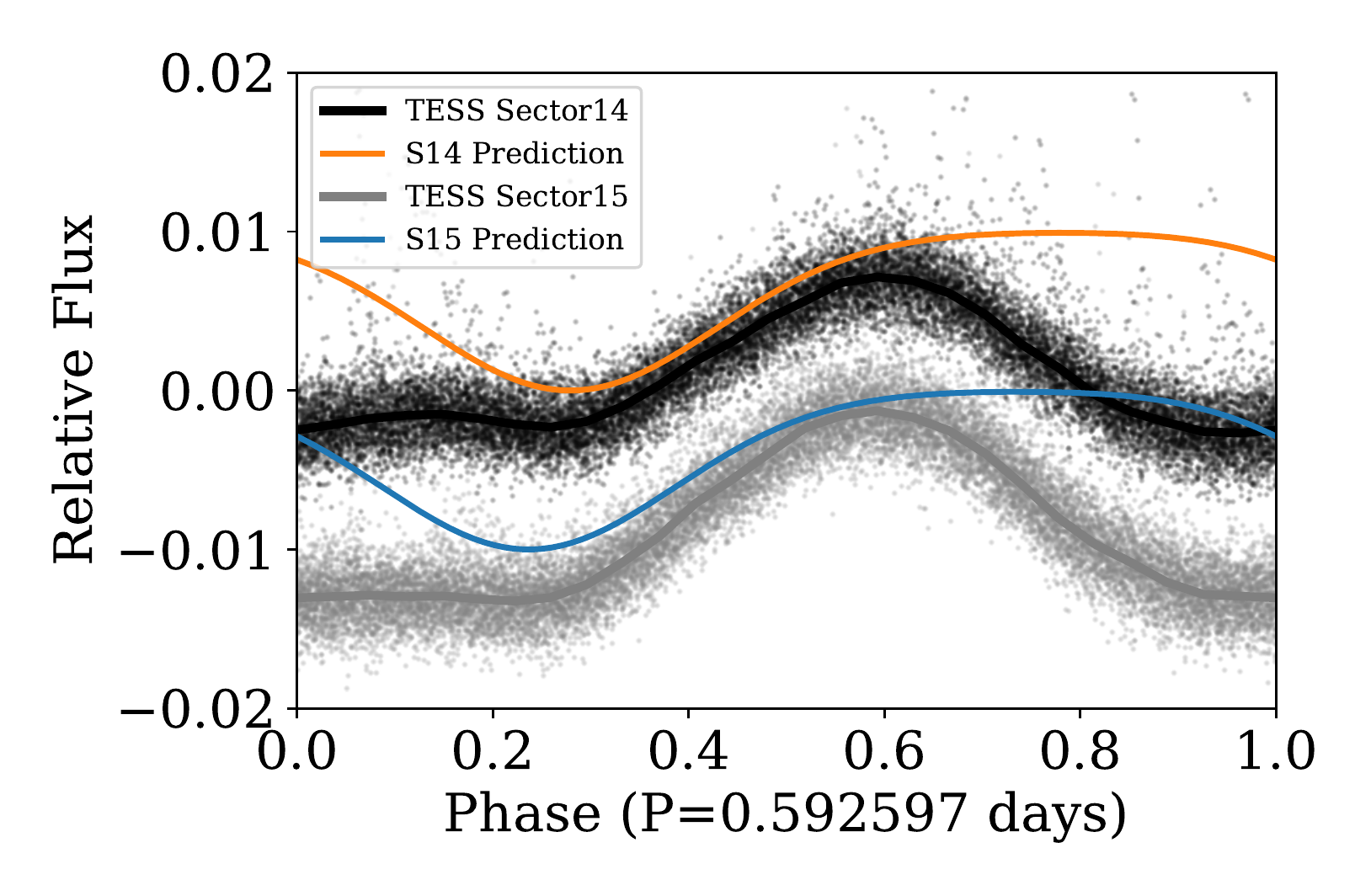}
\caption{
Phase-folded 2-minute cadence light curve TESS Sector 14 (black points) and Sector 15 (grey points) data for GJ 1243. The primary starspot from \citet{davenport2015b} at Phase=0 appears unchanged. The secondary starspot light curve, here modeled as Gaussian curves (orange and blue lines) in TESS Sectors 14 and 15 appears to track the phase location and evolution propagated from the \Kepler data noted by \citet{davenport2015b}.
}
\label{fig:pred}
\end{figure}

We then propagated this updated differential rotation rate from \Kepler to the new TESS observations, and find the secondary spot is at a phase consistent with the predicted phase. In Figure \ref{fig:pred} we also find the the secondary spot appears to move slightly in phase between TESS Sector 14 and Sector 15 with the same shear rate seen in \Kepler, while the primary feature remains at a fixed longitude. As seen in \Kepler, the secondary spot in TESS also has approximately the same amplitude in the light curve, indicating these starspots (or starspot groups) have roughly equal surface filling factors (areas).

The consistent location and evolution in phase of the secondary starspot seen in TESS as compared with \Kepler has two possible interpretations: 1) this feature is the same starspot observed in the latter years of the \Kepler mission, which has not changed significantly in latitude (i.e. change the differential rotation rate) or size relative to the primary spot. 2) The secondary spot seen in TESS is a newer feature than the one observed with \Kepler and is coincidentally aligned with the predicted phase, but is likely at the same latitude since it exhibits the same differential rotation rate. 
Since we lack monitoring of GJ 1243 throughout the 6-year gap between the \Kepler and TESS missions, it is impossible to show that a secondary starspot has remained consistent in its phase evolution throughout that time. As the secondary spot seen in the first 300 days of \Kepler also does not appear to be in sync with this phase evolution, we prefer the latter interpretation, that the secondary starspot seen in TESS is coincidentally in phase with the feature found in \Kepler.

\section{Discussion}
\label{sec:discuss}

Using a combination of data from the \Kepler and TESS missions we have analyzed space-based precision photometry spanning a 10-year baseline for the benchmark active M dwarf, GJ 1243. Our study of flaring and starspot behavior in these light curves sheds new light on the active lives of low-mass stars, and opens the door to searching for slow changes in flare rates due to activity cycles, or surface magnetic field topology via starspot modulations. 

From the new sample of 133 flares from the 2-minute light curves in TESS sectors 14 and 15, we found the flare rate for GJ 1243 has not appreciably changed over the decade timescale. The raw number of flares detected per day is lower in TESS compared to \Kepler, but this is simply due to the lower signal-to-noise in TESS, while the star's actual activity level was unchanged. 
The signal-to-noise of the \Kepler 1-minute cadence light curve is $\sim$6x higher than that of the TESS 2-minute data for GJ 1243. For comparison, the maximum rate of flaring recovered (i.e. for the smallest energy events shown in Figure \ref{fig:ffd}) is $\sim$8x higher in \Kepler than in TESS, which is well aligned with the expected difference in flare yields between the surveys.

As \citet{davenport2019} notes, the Flare Frequency Distribution (FFD) is the most robust parameter space to compare flare activity levels between different stars or between observations of a given star with different noise levels or observing duration. The flare census for GJ 1243 shows no significant variation in either the gross flare rate, or the FFD slope, indicating the same physical process is at work. We also find that flare energies computed using the \Kepler and TESS filters are comparable, making FFDs comparisons between these missions a prime tool in searching for flare rate variations.

The light curve for GJ 1243 shows a two-starspot morphology in both the \Kepler and TESS data, with a rotation period and primary starspot feature that has remained unchanged over 10 years. 
The nearby, rapidly rotating, flaring M4 dwarf V374 Peg is the most natural comparison star for our long-term space-based starspot monitoring of GJ 1243. Over 16 years of ground-based monitoring, \citet{vida2016} found the two starspots for V374 Peg have remained effectively stationary in longitude and with constant amplitude (area coverage). This starspot stability is consistent with previous measurements of a strong dipolar magnetic field from Zeeman-Doppler imaging for V374 Peg \citep{morin2008a}. 
The long term stability of the primary starspot in the new TESS data for GJ 1243 lends support to the interpretation that this feature represents a large polar spot ``cap'', possibly due to strongly dipolar magnetic field that is slightly misaligned with the rotation axis.

The secondary starspot for GJ 1243 has once again been observed to have weak differential rotation. While this secondary spot in TESS is perhaps coincidentally in phase with the evolution seen in \Kepler, the 10-year spot map (Figure \ref{fig:map}) indicates the surface is dominated by two very large, and very slowly evolving features. The presence of two such large and stable active regions poses interesting questions about the surface topology of the magnetic field for GJ 1243. We believe this benchmark star is a prime candidate for continued photometric monitoring, as well as follow-up spectro-polarimetric observations to constrain the surface magnetic field structure as was done for V374 Peg \citep{morin2008a} and other late-type M dwarfs like GJ 1245B \citep{morin2010}.

\citet{hawley2014} found with \Kepler that GJ 1243 showed now correlation between rotation phase and the flare rate or flare energy. This was interpreted as flares occurring in active regions across the entire star, rather than concentrated near the starspots. Using the TESS flare data we again find no indication of rotational phase correlation with flare occurrence or energy. By contrast, \citet{roettenbacher2018} have recently found that small flares {\it do} seem to occur coincidentally with major starspot groups for predominantly higher mass flare stars (G through M) in \Kepler. They interpret this difference as a fundamental change in the surface magnetic topology, in agreement with dynamo models for convective stars \citep{yadav2015} that predict active regions (i.e. flares) are spread across the star's surface.

The constant flare rate and starspot properties for GJ 1243 broadly indicate that the star either lacks a solar-like activity cycle, or the timescale for such a cycle is much longer than 10 years. The latter interpretation is in conflict with the traditional view that long activity cycles correlate with slow rotation \citep{bohm-vitense2007}. However, a lack of an activity cycle is not surprising given the possible young age for GJ 1243 \citep{silverberg2016}, and that the star is rapidly rotating and well within the ``saturated'' dynamo activity regime where cycle behavior is not expected \citep[e.g. see][]{testa2015}. GJ 1243 is a fully convective star, and it is also unknown if Solar-like activity cycle behavior can arise in stars across the fully convective boundary. \citet{yadav2016} have shown simulations that generate activity cycle behavior for fully convective stars, but only those that are {\it slowly} rotating. Recently \citet{bustos2019} found a candidate for such behavior using long term spectroscopic monitoring for the slowly rotating ($\sim$109 day) M4 dwarf, GL 447A.
Long term flare monitoring for all active stars like GJ 1243, including V374 Peg (TIC 283410775, observed in Sector 15), is now possible with TESS. As \citet{scoggins2019} notes, this continuing flare rate census on decades timescales will be a valuable metric for detecting activity cycles for many stars.

\section{Summary}
\label{sec:sum}
We have presented new flare rates and starspot tracing for the active M dwarf GJ 1243 from the TESS mission. Comparing with similar data from the \Kepler mission allows an analysis of magnetic activity on a decade timescale. Both the flaring and starspot behavior appear unchanged, indicating no sign of an activity cycle. The primary starspot is found to be stable in phase, while the secondary starspot feature continues to show signs of weak, solar-like differential rotation.

We have found the flare rates and flare frequency distributions of M dwarfs between \Kepler and TESS are easily compared, despite the significantly longer stare time and larger aperture of the former. The wider bandpass of the TESS filter results in a slightly larger ($\sim$2.4\%) observed energy for a given flare as compared to \Kepler -- well within the nominal error budget of most flare energy estimates.

\acknowledgments

The authors would like to express our sincere gratitude to the anonymous referee for their helpful suggestions that improved this manuscript. 

JRAD acknowledges support from the DIRAC Institute in the Department of Astronomy at the University of Washington. The DIRAC Institute is supported through generous gifts from the Charles and Lisa Simonyi Fund for Arts and Sciences, and the Washington Research Foundation.

GTM acknowledges support from the National Science Foundation Graduate Research Fellowship Program under Grant No. DGE-1762114. Any opinions, findings, and conclusions or recommendations expressed in this material are those of the author(s) and do not necessarily reflect the views of the National Science Foundation.

This research was supported by the National Aeronautics and Space Administration (NASA) under grant number 80NSSC19K0375 from the TESS Cycle 1 Guest Investigator Program, 
and grant number 80NSSC18K1660 issued through the NNH17ZDA001N Astrophysics Data Analysis Program (ADAP).

\software{Python, IPython \citep{ipython}, 
NumPy \citep{numpy}, 
Matplotlib \citep{matplotlib}, 
SciPy \citep{scipy}, 
Pandas \citep{pandas}, 
Astropy \citep{astropy},
Lightkurve \citep{lightkurve}}

\end{document}